\documentclass[sigconf]{acmart}
\usepackage{graphicx}
\usepackage{epstopdf}
\usepackage{graphics}
\usepackage{url}

\AtBeginDocument{%
  \providecommand\BibTeX{{%
    \normalfont B\kern-0.5em{\scshape i\kern-0.25em b}\kern-0.8em\TeX}}}

 \acmConference[Arxiv]{Arxiv Preprint}

\begin{document}

\title{Machine Learning Fund Categorizations}


\author{Dhagash Mehta}
\email{dhagashbmehta@gmail.com}
\affiliation{%
  \institution{The Vanguard Group}
  \city{Malvern}
  \state{PA}
  \country{USA}
}

\author{Dhruv Desai}
\email{dhruv_desai@vanguard.com}
\affiliation{%
  \institution{The Vanguard Group}
  \city{Malvern}
  \state{PA}
  \country{USA}
}

\author{Jithin Pradeep}
\email{jithin_pradeep@vanguard.com}
\affiliation{%
  \institution{The Vanguard Group}
  \city{Malvern}
  \state{PA}
  \country{USA}
}

\renewcommand{\shortauthors}{Mehta et al.}

\begin{abstract}
  Given the surge in popularity of mutual funds (including exchange-traded funds (ETFs)) as a diversified financial investment, a vast variety of mutual funds from various investment management firms and diversification strategies have become available in the market. Identifying similar mutual funds among such a wide landscape of mutual funds has become more important than ever because of many applications ranging from sales and marketing to portfolio replication, portfolio diversification and tax loss harvesting. The current best method is data-vendor provided categorization which usually relies on curation by human experts with the help of available data. In this work, we establish that an industry wide well-regarded categorization system is learnable using machine learning and largely reproducible, and in turn constructing a truly data-driven categorization. We discuss the intellectual challenges in learning this man-made system, our results and their implications. 
\end{abstract}




\maketitle

\section{Introduction}
The problem of identifying similar products from a given universe of products arises in many areas of the business, e.g., recommending similar movies to the ones the customer watched, replacing items on the shelf with similar items in case of understocking, recommending complementing items to the ones the customer is buying based on what other similar customers bought together, etc. \cite{resnick1997recommender,aggarwal2016recommender}. Even though in general the similarity problems arise with clear statement and have impactful applications, quantifying similarity usually turns out to be a highly challenging problem: removing emotional aspects as well as biases from the process, identifying a metric of similarity, identifying the features that determine similarity, and determining potential nonlinear relationship among the features are some of the challenges in any similarity related problems. 

In the financial domain, the product similarity may be phrased in terms of identifying similar financial entities such as stocks, bonds, funds, etc. In the present work, we focus on similarity between funds (which include Exchange Traded Funds (ETFs) for the purpose of this paper). Such a similarity computation can then be employed for multiple business applications such as in recommending a product against a competitor's fund; to explain similarities and advantages of home-grown products compared to competitors' products for marketing purposes; to perform tax loss harvesting at the end of the tax season; to quantitatively measure portfolio diversification in a portfolio, etc. Industry wide fund ratings also first identify peer groups built upon such categorizations, and then rate them according to certain criteria.

One of the most popular approaches to find similar funds is a categorization of funds provided by data vendors such as Morningstar \cite{morningstarglobalcategory} and Lipper \cite{lippercategory}. Such categorizations are a result of recommendations of committees of experts based on several quantitative and qualitative assessments \cite{morningstarequitysectorbreakdown}, and are generally well regarded in the industry.

However, many a times such a third-party categorization may depend on the descriptions in the fund brochures supplied by the fund managers than on a purely quantitative practice. Any qualitative portion of the decision making process is also likely to add unintentional biases as well as emotional aspects. In \cite{marathe1999categorizing} (see, also, e.g., \cite{orphanides1996compensation,brown1997mutual,dibartolomeo1997mutual,elton2003incentive}), it was argued that there are multiple ways the commercially available categorizations may be tricked to misclassify a mutual fund to a different category: the mutual fund managers may have incentives to get their managed funds to be compared with wrong peers in order to make the fund appear better in comparison, and hence they may provide selective information to the categorization committee; the incentive structure of the fund managers may provide further reasons to enforce a misclassification; qualitative nature of some of the analysis that may go into the categorization process may trigger unintentional misclassifications; etc. In \cite{marathe1999categorizing}, then, a purely data-driven (clustering) approach to categorize the then available mutual funds was proposed, and results were compared to the contemporary Morningstar categorization. This, and many other studies \cite{kim2000mutual,castellanos2005spanish,moreno2006self,acharya2007classifying,haslem2001morningstar,lamponi2015data} found "inconsistencies" between a typical third-party categorization and the results of their clustering analysis. Below, we review the previous research on data-driven categorizations and discuss their limitations.

\subsection{Previous Approaches for Data-Driven Categorization}
The existing literature on data-driven categorization has been focused on unsupervised clustering of funds. Unsupervised clustering \cite{cai2016clustering} is a conventional data-driven technique where a clustering algorithm identifies clusters of similar entities in a high-dimensional space of the pre-selected variables. Here, the ''similarity'' can be defined in terms of Euclidean distance, Jaccard distance, cosine of the angle between vectors of mutual funds, etc. The clustering techniques and the aforementioned similarity computations are two sides of the same coin as they both identify similar entities (the former lumps them together in corresponding clusters). The main difference between a traditional similarity computation and unsupervised clustering approaches boils down to the process of determining the centroids of the clusters: the latter lets the algorithm determine the centroids, whereas in the former the user hand-picks the ''centroid''.

One of the first unsupervised clustering of mutual funds was performed in \cite{marathe1999categorizing} (see \cite{mcdonald1974objectives,martin1982fund,dibartolomeo1997mutual,sharpe1992asset} for earlier attempts of classifying mutual funds specially for return based and investment objective based methodologies) using the then available Morningstar variabls, and observed that \textit{with respect to their clustering} (over their 28 preselected variables: Morningstar risk (3 year and 5 year), Morningstar returns (3 year and 5 year), 1 year total return, annualized return (3 and 5 year), annual return (last 4 years), 3 year alpha and beta, standard deviation (3 year and 5 year), income ratio, turnover, potential gain exposure, \% cash, \% stocks, \% bonds, \% preferred, \% other, maximum sales charge, \% front load, \% deferred, \% expense ratio, net assets), the Morningstar categorization misclassified 43\% mutual funds. 

In \cite{haslem2001morningstar}, all the then available US large-cap mutual funds were clustered into 3 clusters using the k-means clustering technique with Euclidean distance as the similarity metric, with 11 preselected variables: cash ratio (\%), P/E and P/B ratios, 3-years earning growth (\%), median market cap (\$), turnover (\%), foreign (\%), stocks (\%), asset in top 10 (\%), number of securities and top 3 sectors (\%). They showed that 38 out of 83 funds were misclassified by Morningstar with respect to their clustering. 

In \cite{lamponi2015data}, a handful of investable assets (cash, equities, fixed income, real assets and alternative investments) were clustered into 5 clusters using a hierarchical clustering approach to yield that some of the investable assets fell into clusters of another types of assets (e.g., some real assets were classified as fixed income) when applied purely data-driven clustering than their categories.

Authors in \cite{lajbcygier2008soft} argued that boundaries between different investment styles are continuous rather than "hard", and then used a soft clustering technique called Fuzzy-C-Means to demonstrate that such a technique can predict mutual fund performance better out-of-sample than a hard clustering technique. In \cite{vozlyublennaia2018mutual}, clustering approach was used to identify \textit{unique} funds (which create "clusters" only of their own) and concluded that the unique funds have higher total expense ratios mainly because of their specialized management styles and in turn higher management fees.

In \cite{menardi2015double}, a two stage clustering was performed for 1436 funds using 24 variables (expected excess return, tail gain, expected tail gain, Jensen's alpha, standard deviation of returns, lower partial moment, value at risk, expected tail loss, maximum drow-down, Treynor index, appraisal ratio, Sharpe ratio, Sortino ratio, Kappa 3 index, upside potential ratio, Omega index, Farinelli-Tibiletti index, Burke ratio and Sterling ratio): first, the variables themselves are clustered to obtain groups of variables which are homogeneous with respect to the information they explain, then each group of variables is dimensionally reduced using Principal Component Analysis (PCA), and finally a hierarchical agglomerative clustering is performed on the reduced dimensional space. This work comes closer to the philosophy of the present work in terms of importance of finding the most crucial features, though a PCA only captures linear relationship among variables whereas we aim to capture nonlinear relationship when reducing the dimensionality.

In \cite{moreno2006self} a nonlinear clustering technique called self-organizing map (SOM) \cite{kohonen1990self} was used to cluster 1592 mutual funds from the Spanish market over ten fund attributes (average return, standard deviation, skewness, kurtosis, the 5\% of maximum losses, the 5\% of maximum returns, the reward-to-semivariability ratio, the beta against a chosen index, the beta against a 10-year notional bond and the correlation of each fund with an equally weighted benchmark obtained from each of the 14 legal categories in the Spanish market). They used an artificial neural network based SOM which configures the output units into a topological representation of the original data. Here, the prototype vectors are positioned on a regular low-dimensional (typically, 2-dimensional) grid in an ordered fashion. Due to the low-dimensional grid representation, SOMs may also be used for visualization of the clustered data. The authors concluded that the clustering using SOMs had significantly less missclassifications with respect to the categories than the one using KNN.

In \cite{sakakibara2015clustering}, a network of 551 Japanese mutual funds was constructed using top 10 stocks of each fund. Here, the bipartite network was projected to a unimode network of funds where the weights between nodes were the number of common stocks. Then, the weighted network was clustered using the k-means and spectral clustering method for graph partitioning. Eventually, the clusters were compared with Morningstar categories and a few qualitative differences between the two were observed.

In \cite{pattarin2004clustering,agudo2005does,lytkin2008variance,corduas2008time,lisi2010clustering,gerlach2017stable}, clustering of mutual funds was performed based on daily returns and prices rather than investment styles. The reader is referred to \cite{das2003hedge,miceli2004ultrametricity,baghai2005consistency,gibson2007style,shawky2010stylistic} for studies on clustering of hedge funds which is out of scope for the present study.

\subsection{Our Contribution}
We first emphasize that a mismatch between the data-driven approaches to categorize the mutual funds and Morningstar categorization of the funds does not mean one approach is better or worse than the other, as they all have different purposes and underlying methodologies \cite{haslem2001morningstar}. 

Our starting point is the assertion in \cite{haslem2001morningstar} (see also comments by Gambera, Rekenthaler and Xia in \cite{haslem2001morningstar}) that such categorizations as the Morningstar categorization are a result of a \textit{classification system} devised by a committee of experts rather than clustering: the categorization system is based on the extensive (sometimes proprietary) data available to the panel as well as the expertise and experience of the panel. For the investors who are in search for similar funds with respect to these variables, the Morningstar (and Lipper, for that matter) categories provides an easy look-up table.

In the present work, unlike the previous works reviewed above, we take a bottom-up approach which is to learn the data-categorization as a supervised classification problem where we attempt to learn a Morningstar categorization given various characteristics of the funds provided by Morningstar Direct. Out approach has multi-fold advantages: first, our approach validates the hypothesis that an expert committee provided categorization is learnable using a machine learning algorithm, and hence is reproducible. Moreover, we only use the aggregate level holding information for the funds, unlike the previous research which used either performance related variables or a combination between holding and performance information. Once trained, the machine then can also be employed to classify a completely new fund to closely mimic the committee. We can then also identify the important features and their weights learned by the machine.

\section{Data and Problem Definition}
For the scope of this problem we used Morningstar Direct as the primary data source. This data was for the month of April 2020. This dataset contains all US domiciled ETFs, money market funds and open-end mutual funds. We selected only the oldest share class funds i.e., fund in a given share class with longest history to preserve distinct funds out of the same shared classes. Eventually, we had a total of 2352 ETFs, 364 money market funds and 7601 open-end mutual funds, i.e., a total of 10,318 funds. 

\begin{figure}[h]
    \centering
    \includegraphics[width=0.45\textwidth]{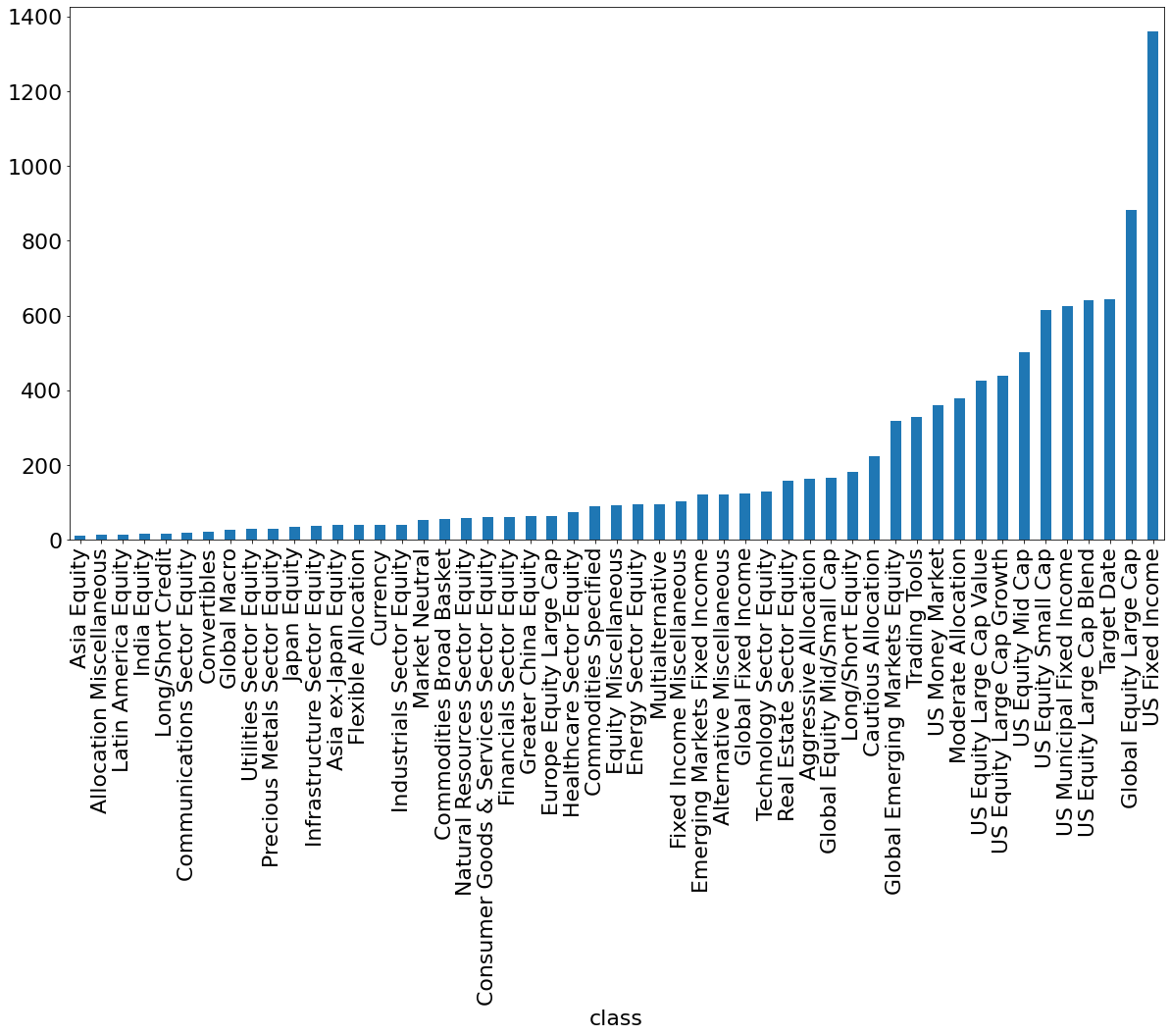}
    \caption{Distribution of the available 10,300 funds within 51 categories.}
    \label{fig:category_bar_graph}
\end{figure}

Morningstar has four types of categorizations each having more granularity and hence more categories: Broad categories, Global categories, Morningstar categories and Morningstar Institutional categories. For the purpose of this work, we chose the Global categories as they were granular enough to provide a rich set of sub-problems, whereas coarse enough to give a better chance to be learnable using the limited variables available to us. There were a total of 61 Morningstar Global categories \cite{morningstarglobalcategory} for the 10,318 funds we sorted out from the data. In order to maintain balanced class weight to train the algorithms we chose a threshold of minimum three examples per categories, which finally left us with a total of 51 categories and 10,300 funds. The funds are not uniformly distributed among all the 51 categories, and the distribution is shown in Figure \ref{fig:category_bar_graph}.

Table \ref{table:feature_list} contains the entire list of variables which were treated as input features. For Type in Table \ref{table:feature_list} listed as numeric are percentage value for that feature for the given fund. Numeric features in this dataset were present for both long and short attributes of the fund. We included all the long features and excluded the short attributes due to its sparsity.

Some features included in this dataset are aggregations of other features present in the dataset. In order to capture the entire dataset the lowest feature in the hierarchy were chosen. For equity features they follow a hierarchy of Equity Econ Super Sectors which are further divided into eleven Equity Econ Sectors. These sectors are broken down to 55 Industry groups. These industry groups are further classified into to 145 Equity Industry types \cite{morningstarequitysectorbreakdown}. For the Fixed Income features, Fixed Income Super Sectors are broken down into Fixed Income Primary Sectors which are further broken down into Fixed Income Secondary Sectors. We included the Fixed Income secondary sector in the training set as the remaining two features can be calculated by grouping these features \cite{morningstarfixedincomeclassificationreakdown}. The Holdings features (number of Stock Holdings (Long), number of Bond Holdings (Long) and number of Other Holdings (Long)) are the actual number of holdings in the given type (Stocks, Bonds and other) which are held by the given fund.

Some features included in this dataset are correlated with other features, e.g., Equity Region which gives a region wise break down of the percentage of Equity in a specific region for the given fund is correlated to Equity Country. Equity Country is the country wise breakdown of the percentage of equity of the fund in a country. Since Regions are made up of country these two features are highly correlated but provide different flavor of the aggregation. We keep these features in our computation, and let the algorithms learn their interdependencies and importance on their own. \\
\begin{table}[h!]
\begin{tabular}{l|l|l}
\hline
\textbf{Feature name}                  & \textbf{Total features} & \textbf{Type} \\ \hline
Asset Allocation                       & 8      & Numeric       \\ \hline
Benchmark                              & 2      & Categorical   \\ \hline
Coupons                                &4       & Numeric       \\ \hline
Calculated Credit Rating               &7       & Numeric       \\ \hline
Equity Industry                        & 148    & Numeric       \\ \hline
Equity Country                         & 53     & Numeric       \\ \hline
Equity Region                          &16      & Numeric       \\ \hline
Equity Style                           &15      & Numeric       \\ \hline
Fixed-Inc Secondary Sector             & 183    & Numeric       \\ \hline
Fixed-Inc Country                      & 54     & Numeric       \\ \hline
Fixed-Inc Region                       & 3      & Numeric       \\ \hline
Fixed-Inc Sector Government            & 68     & Numeric       \\ \hline
Holdings                               & 4      & Numeric       \\ \hline
Market Cap                             & 12     & Numeric       \\ \hline
Maturity                               &13      & Numeric       \\ \hline
MSCI Country                           & 156    & Numeric       \\ \hline
Muni                                   & 59     & Numeric       \\ \hline
Product Involvement \%                 & 15     & Numeric       \\ \hline

\end{tabular}
\caption{Variables list. The numeric values are always percentages, and categorical values are names of different benchmarks.}
\label{table:feature_list}
\end{table}

Now, we formulate the problem of learning categorization as a machine learning problem: given all the input features as given in Table \ref{table:feature_list}, learn to classify each fund into one of the 51 categories (Figure \ref{fig:category_bar_graph}). In other words, we reformulated the data-driven categorization problem as a multi-class classification problem.

\section{Methodology}
In this section, we describe our methodology. In particular, we first briefly describe the algorithms employed for the problem at hand, and then explain the details of our modelling process. 
\subsection{Models}
Since we formulated the problem of learning Morningstar categorization as a supervised multi-class classification problem, we had a large pool of algorithms to chose from. With some exploratory analysis, we determined three different models most suitable for our purposes: Decision Tree, Random Forest and Deep Artificial Neural Network. This ordering of the list of models also represents increase in complexity and overall efficiency, and decrease in interpretability.

\subsubsection{Decision Tree}
Decision Tree is one of the simplest yet powerful machine learning algorithms for both classification and regression tasks \cite{hastie2009elements}. The algorithm attempts to identify the decision process for the classification task from the given data: one first constructs a cost function, which in the classification task could be Gini or information entropy. The algorithm then starts by considering all the features first, and tries different split points for each feature to identify the most cost efficient (minimum cost) split. The feature for the most cost efficient split is then the top most level of the decision tree. For each of the branches of the split at this level, one iterates the same splitting process for the remaining features. The algorithm stops when the predetermined number of levels (called the depth) is reached. The depth is a hyperparameter and can be tuned to control the learning process. Although the results from Decision Tree are easy to interpret, it is prone to overfitting.

\subsubsection{Random Forest}
Random Forest \cite{hastie2009elements, liaw2002classification, breiman2001random} is an ensemble learning method again based on Decision Trees. Here, instead of constructing one decision tree, one constructs multiple decision trees only based on randomly selected subsets of features (alternatively, randomization may also be applied on data). Then, an aggregation of all these smaller trees is used to obtain a final model, hence the name ''Random Forest''. The depth of each decision tree and the number of decision trees both are hyperparameters and can be tuned to improve learning. Random Forest evades the overfitting problem of an individual decision tree by the aggregation (i.e., ensembling), making it more accurate but less interpretable due to the presence of multiple small trees. However, one can extract feature importance from this model, hence at least retaining an important portion of interpretability which we will utilize in present work.

\subsubsection{Deep Artificial Neural Network}
Deep artificial neural networks, or deep learning techniques \cite{cybenko1989approximation,hornik1989multilayer, lecun2015deep, bengio2015deep}, have recently become very popular due to their success in computer vision tasks such as image recognition, scene understanding, and object detection, as well as in many other applications such as natural language processing, recommender systems, speech recognition, etc. In a nutshell, the so-called feedforward deep networks, i.e., a multi-layer structure of compositions of nonlinear functions or activation functions, have been shown to be universal approximators for any target function as long as the network parameters and the multi-layer structure are chosen carefully~\cite{cybenko1989approximation,hornik1989multilayer}. The number of layers of compositions of nonlinear functions is called the depth of the network. One can resort to  hyperparameter optimization to find an appropriate architecture of the network, and to stochastic gradient descent or one of its variants to find appropriate values of the network parameters \cite{bengio2015deep}. Deep networks are empirically shown to learn the data with high accuracy and generalize well, however, are also highly opaque when it comes to interpretation of the results. 

For the deep neural networks, we used a feed-forward neural network that consist of 3 fully connected hidden layers having 512, 256 and 128 hidden units, respectively, with rectified linear unit (ReLU) activation function. The output layer applies softmax activation, ensuring output values are in the range $[0,1]$ over the class distribution. The network is trained using the Adam (Adaptive moment estimation) \cite{kingma2014adam} optimizer with categorical cross-entropy loss function.

\subsection{Data Pre-processing} 
Before employing the above models, we perform pre-processing of the data as follows: all numeric percentage features in this dataset were rounded to 4 decimals. All the missing values were replaced with zero (there was no zero value in these features prior to imputation), since numeric percentage features represents the percentage of fund invested in that attribute. Then, the zero percent would then indicate that the fund has no investment in the given attribute type, which is the correct interpretation of the missing values in the given data (unless for a genuinely corrupt data-point which we may not be not aware of). For the Holdings features, missing values were also replaced with zero with the same reasoning. We used MinMax scaling on the entire training matrix to have all feature values between zero and one. For a given feature MinMax scaled value can be written as $X_{scaled}=(X - X_{min})/(X_{max}-X_{min})$, where $X_{min}$ and $X_{max}$ are the minimum and maximum value for the entire feature, respectively \cite{scikit-learn}.

\subsubsection{Feature Engineering} The two benchmark features available in the dataset, namely, S\&P Dow Jones Benchmark and FTSE/Russell Benchmark, are converted to one-hot encoded vectors. There are a total 89 unique values for S\&P Dow Jones Benchmark and 42 unique values for FTSE/Russell Benchmark. For the remaining features in Table \ref{table:feature_list} are percentage values of the given fund present in that attribute. The final data matrix had the dimension of $10,300 \times 875$, where columns are composed of both categorical and numeric features.


\subsubsection{Training-testing split (stratified)} We used a training-testing split of $75\%$ and $25\%$, respectively. We used stratified split: as the classes are imbalanced, stratification ensures that train and test split approximately have the same percentage of samples of each target class as the complete set.

\subsection{Modelling Details}
Here, we describe details on the modelling methodology.

\subsubsection{Balanced Class weight} We used balanced the class weight technique to train the Decision Tree and Random Forest algorithms to handle the class imbalance present in the dataset. In these cases, the class-weights are inversely proportional to their frequencies. The weight assigned to each class is then $w_i=\frac{n}{C n_i},$
where $w_i$ is the weight for class $i$, $n$ is the number of observations, $n_i$ is the number of observations in class $i$ and $C$ is the total number of classes.

\subsubsection{Cross validation} Learning the parameters of a prediction function and testing the model on the same data causes over fitting. In order to perform hyperparameter tuning and ensure that the test set does not leak into the model and evaluation metrics, we used cross validation for Decision Tree and Random Forest: we used the $k$-fold cross validation method where the training set is spilt into $k$ smaller sets. The model is trained on $k -1$ of the folds as training data and validated on the remaining part of the set. We used stratified $k$-fold cross validation which ensured that all the folds the same percentage of samples of each target class. We used 5-fold cross validation in our work.

The same train-test split was used for the deep neural network with holdout cross validation to evaluate the model performance on validation set($10\%$ of the training dataset).

\subsection{Metrics}
Because we had a highly imbalanced dataset at hand, we used different metrics to measure the accuracy of our models.
\subsubsection{Accuracy} 
Accuracy can be defined as the fraction of predictions which were predicted correctly by the model. For multi-class classification problems, accuracy can be calculated over the entire set by assigning the subset accuracy of $1.0$ if the samples matches the true label set, else zero. To be sensitive to the performance of individual classes, we used the weighted accuracy: weight $w_i$ is assigned to every class, such that $\sum_{i=1}^{C} w_i = 1$. The higher the value of $w_i$ for a given class, the greater is the influence of the given class on the weighted accuracy. If $\hat{y}_{i}$ is the predicted value of $i^{th}$ sample and $y_i$ is the corresponding true values, then the weighted accuracy over $n$, the number of data-points, is defined as \cite{scikit-learn}
\begin{equation}
\notag
accuracy(y,\hat{y}) = \sum_{j=1}^{C} w_j \sum_{i=1}^{n} 1(\hat{y}_i=y_i),
\end{equation}
where 1(x) is the indicator function.

\subsubsection{F1 score}
The F1 score is defined as below:
\begin{equation}
\notag
    F1\,Score = \frac{2(Precision \times Recall)}{Precision +\,Recall},
\end{equation}
where 'Precision' is defined as $TP/(TP + FP)$ and 'Recall' is defined as $TP/(TP+FN)$, with TP being the number of true positives, FP being the number of false positives and FN being the number of false negatives. However, the F1 score would not directly apply to multi-class classification problems with imbalanced data such as ours. Instead, we used the micro F1 score, which favors all the classes equally, and the macro F1 score which calculates the F1 score for each label and then take their unweighted sum \cite{scikit-learn}.



\begin{figure}[h]
    \centering
    \includegraphics[width=8cm,height=8cm,keepaspectratio]{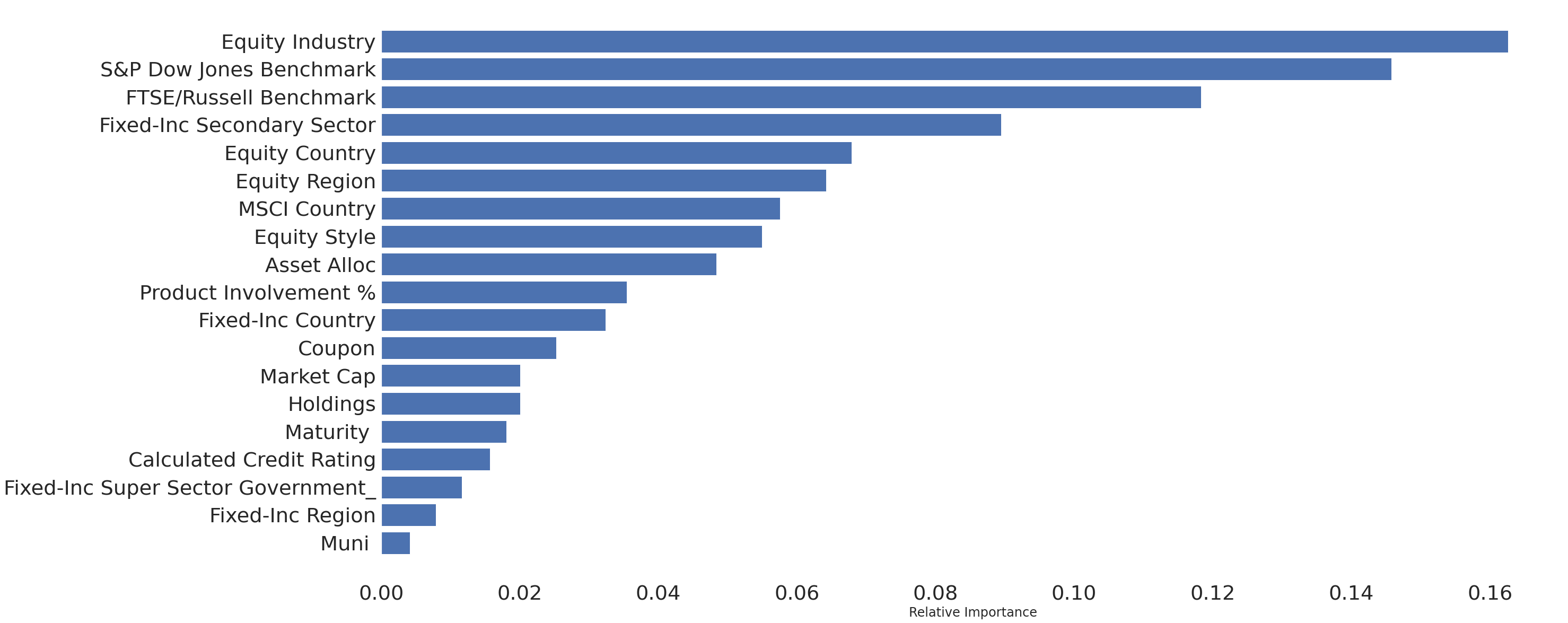}
    \caption{Feature importance from Random Forrest.}
    \label{fig:feature_importance}
\end{figure}

\subsubsection{AUC-ROC}
The Receiver operating characteristic (ROC)\\curve is a plot which illustrates the performance of a binary classifier system as a function of the discrimination threshold. It is created by plotting the fraction of true positive rate (TPR) vs. the fraction of false positive rate (FPR). When using the normalized units, the area under the ROC curve (AUC-ROC) is the probability that a classifier will rank a randomly chosen positive instance higher than a randomly chosen negative instance. For multi-class classificationm this logic can be used by changing it to one vs. rest which is the average of the AUC-ROC for each class against all other, or by using one vs. one and averaging the pairwise AUC-ROC \cite{10.1016/j.patrec.2005.10.010,10.1023/A:1010920819831}. We used micro and macro versions of the AUC-ROC in the present work \cite{scikit-learn}.
\begin{figure}[h]
    \centering
    \includegraphics[width=9cm, height=5cm]{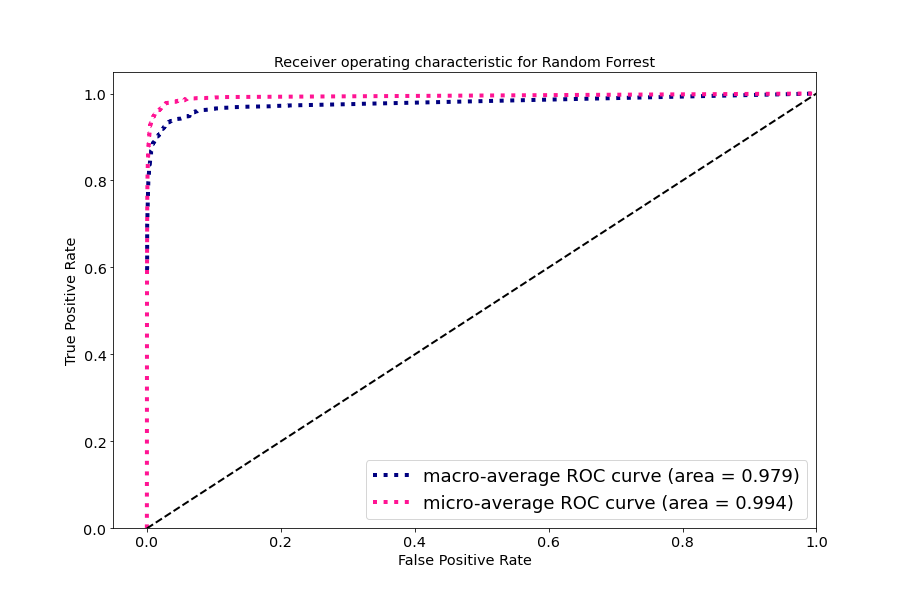}
    \caption{ROC curves for the test data from the Random Forest model.}
    \label{fig:roc_micro_macro}
\end{figure}
\begin{figure*}[h!]
    \centering
    \includegraphics[width=17cm, height=6cm,keepaspectratio]{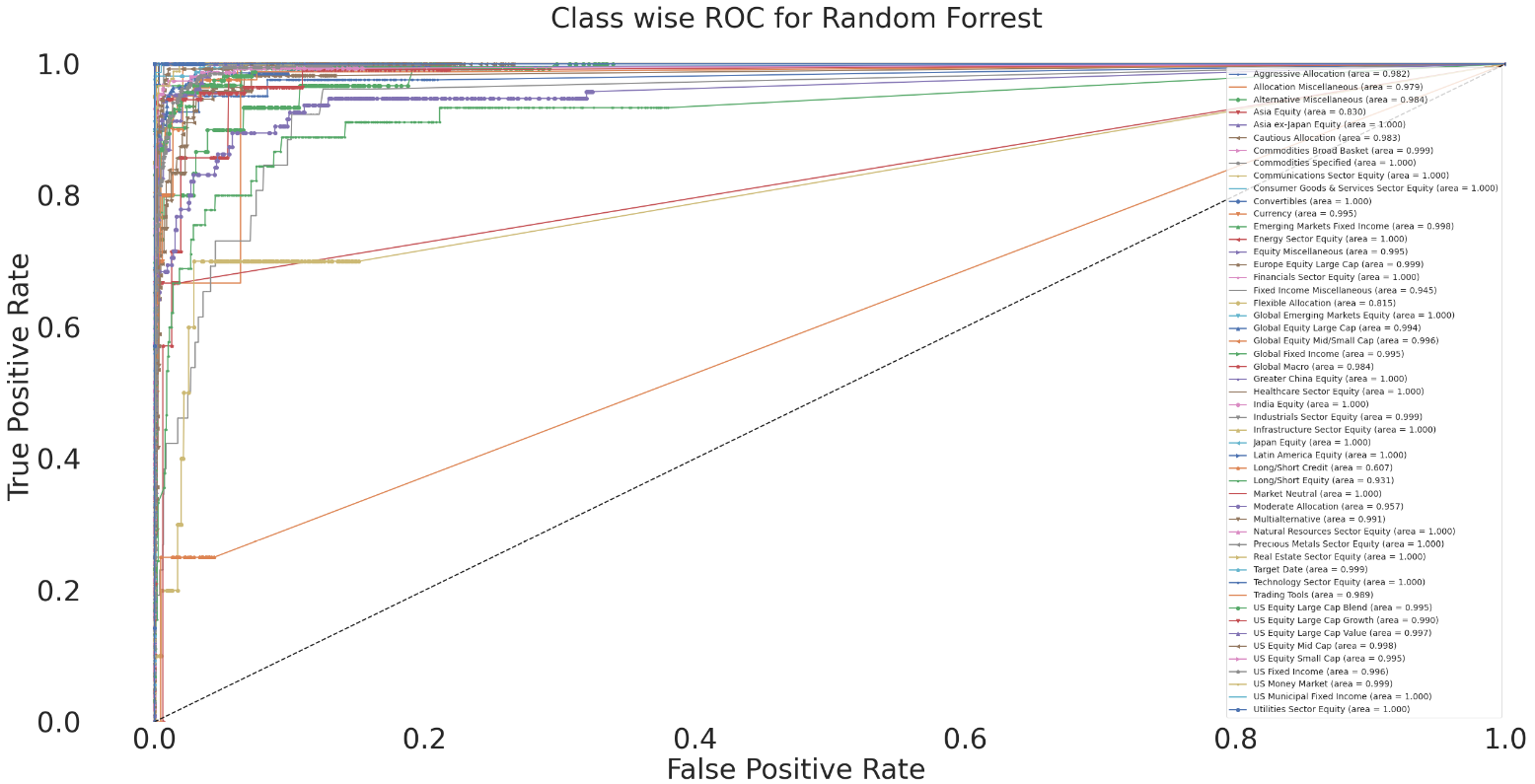}
    \caption{ROC curves for individual classes for the test data from the Random Forest model.}
    \label{fig:roc_all_class}
\end{figure*}
\subsubsection{Confusion Matrix} 
Confusion matrix can be defined as matrix $M$ such that $M_{ij}$ is equal to the number of observations known to be in class $i$ and predicted to be in group $j$ \cite{scikit-learn}. As there are 51 target classes we have a confusion matrix of size $51 \times 51$. We use normalized confusion matrix\cite{scikit-learn} where each count is divided by the sum of each row. The diagonal values ($i=j$) give the true positives for the $i$-th class.

\subsection{Hyperparameter Optimization}
For Random Forrest algorithm we performed a grid search over total number of estimators and criterion parameters. We chose total number of estimators between 80 and 140 with a step size of 20, and used \textit{Gini} and \textit{Entropy} as for criterion. 

\subsection{Computational Details}
These computations were run using a python script with scikit-learn library \cite{scikit-learn} version 0.22.2, and Keras \cite{chollet2015keras} version 2.2.4. We used a standard laptop with 64-bit Windows 10 Operating System, Inter(R) Core(TM) i5-8350u CPU @ 1.70 GHz and 16.0 GB of RAM.
\begin{figure*}[]
    \centering
    \includegraphics[width=0.95\textwidth,keepaspectratio]{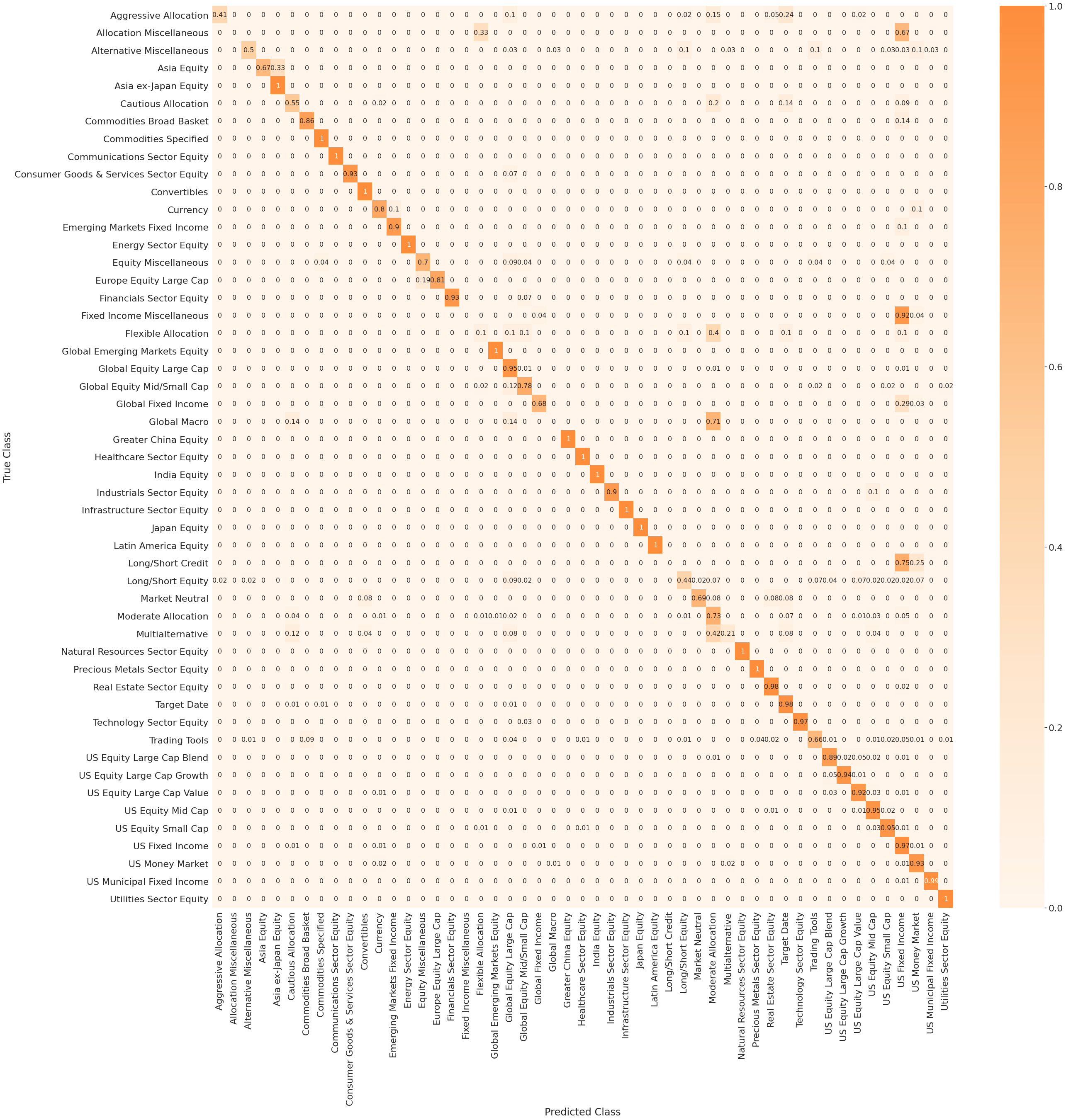}
    \caption{Final confusion matrix on the test data using Random Forest model.}
    \label{fig:conf_mat}
\end{figure*}
\section{Results and Discussion}\label{sec:results_discussion}
Here, we present our results and discuss their interpretations. Table \ref{table:results} summarizes all our results. Though the Decision Tree exhibited relatively lower scores on various metrics compared to the other two algorithms, it provided a solid benchmark for the other algorithms. The deep networks was the best performing model with respect to all the metrics, though Random Forest turned out to be a close second. 

We discuss the results from Random Forest in more detail as the model provided high enough scores on all the metrics, whereas it is also able to provide some further insights on the feature importance. Figure \ref{fig:roc_micro_macro} shows the micro and macro average ROC curves for all the classes. Both micro and macro average AUC-ROC scores are close to 1.0 yielding very high overall accuracy for all the classes in general, meaning our machine(s) has learned the Morningstar categorization system fairly completely. Figure \ref{fig:feature_importance} shows all the features and their importance scores learned by the model. The top three most important features were Equity Industry, S \& P Dow Jones Benchmark and FTSE/Russell Benchmark. These features essentially play a major role in the categorization according to the model. On the other hand, when the three features struggle in reproducing the existing categorizations, the remaining features help resolve the classification. This hierarchy of the features is precisely what is captured by all the models used in this work in their own ways.
\begin{table}[h!]
\centering
\begin{tabular}{l|l|l|l|ll}
 & \textbf{Acc} & \multicolumn{2}{l|}{\textbf{F1-Score}} & \multicolumn{2}{l}{\textbf{AUC-ROC}} \\ \cline{3-6} 
 & \textbf{} & \textbf{Micro} & \textbf{Weighted} & \multicolumn{1}{l|}{\textbf{Micro}} & \textbf{Weighted} \\ \hline
\textbf{DT} & 0.870 & 0.853 & 0.853 & \multicolumn{1}{l|}{0.928} & 0.891 \\
\textbf{RF} & 0.902 & 0.869 & 0.857 & \multicolumn{1}{l|}{0.992} & 0.979 \\
\textbf{DNN} & 0.907 & 0.907 & 0.902  & \multicolumn{1}{l|}{0.996} & 0.987
\end{tabular}
\caption{Final results from all three models. DT, RF, DNN and Acc refer to Decision Tree, Random Forest, DNN and accuracy, respectively.}
\label{table:results}
\end{table}

Figure \ref{fig:roc_all_class} shows the ROC curves for individual classes to further resolve the results from the model. As can be seen from these curves, most categories are perfectly classified by the model, except only a few categories. To investigate the misclassifications in more details, we looked into the confusion matrix and feature importance, in Figure \ref{fig:conf_mat} and \ref{fig:feature_importance}: Most of the misclassifications in fund categories Allocation Miscellaneous, Fixed Income Miscellaneous and Long/Short Credit Categories were to US Fixed Income Category. This behaviour was expected as the S\&P Dow Jones Benchmark feature does not have specific benchmarks that distinguish these categories (the fact is also suggested by 'Miscellaneous' within the names of the categories). Similarly, there are always missing values in the FTSE/Russell Benchmark feature for these specific funds. Both benchmark features are ranked second and third in the feature importance as seen in Figure \ref{fig:feature_importance}. A similar situation arises from Aggressive Allocation, Cautious Allocation, Flexible Allocation, Global Macro and Multi-alternative categories which are often misclassified as Moderate Allocation.
To understand the misclassifications further, instead of assigning the category with highest probability score as the predicted category, we considered top three categories according to the probability scores. For the previously misclassified funds, the ground truth category was almost always in the top three predicted categories.

In most cases, if not all, the fund name itself provides the category name. One would wonder if encoding the fund name using natural language processing technique and using it as a separate variable could improve the classification rate. While this will indeed help in terms of classification accuracy, we emphasize that the funds are named by their fund managers. Fund names lack standardization and this could potentially be misleading and add biases. 


\section{Outlook and Conclusion}
Systematic categorization of mutual funds has become more important than ever because of the wide variety of investment strategies and funds available in the market. The current best categorization systems are provided by third-party data vendors such as Morningstar and Lipper where a committee of experts categorizes a fund as per the available data and their domain expertise and experience, in turn, includes qualitative aspects in addition to quantitative ones. Previous research have questioned these categorizations by pointing to the mismatches between their results on data-driven (unsupervised) clustering techniques with the pre-selected data and the Morningstar categorization.

We note that the fund universe is a man-made system, and whenever investment management firms found ``gaps`` in the high-dimensional space, they tried to fill it with a new fund. Hence, the categorization system provides a unique intellectual challenge to be investigated using machine learning techniques. In this work, we took a more systematic and bottom-up approach to pose the problem as a multi-class (supervised) classification problem. For the first time, we establish that the expert-driven Morningstar categorization at the chosen granularity level is indeed learnable and hence reproducible. The mismatches found between our machine learning system and the Morningstar categorization may be attributed to unavailability of certain variables. The machine then can be used to mimic the committee to categorize any new fund in the market. Any remaining mismatches, even after all the variables are available, may provide interesting cases to be analyzed. Machine learning the more granular categorizations using high-end algorithms will provide a fertile ground of research and potentially novel insights into the funds landscape.


\begin{acks}
We thank Victor Allen, Mirim Lee, Eduardo Fontes and Hussain Zaidi for their insights into the present work. The present work is a product of exploratory research. The views expressed in the paper are solely of the authors and not necessarily of The Vanguard Group.
\end{acks}


\bibliographystyle{ACM-Reference-Format}
\bibliography{sample-base}


\begin{thebibliography}{49}


\ifx \showCODEN    \undefined \def \showCODEN     #1{\unskip}     \fi
\ifx \showDOI      \undefined \def \showDOI       #1{#1}\fi
\ifx \showISBNx    \undefined \def \showISBNx     #1{\unskip}     \fi
\ifx \showISBNxiii \undefined \def \showISBNxiii  #1{\unskip}     \fi
\ifx \showISSN     \undefined \def \showISSN      #1{\unskip}     \fi
\ifx \showLCCN     \undefined \def \showLCCN      #1{\unskip}     \fi
\ifx \shownote     \undefined \def \shownote      #1{#1}          \fi
\ifx \showarticletitle \undefined \def \showarticletitle #1{#1}   \fi
\ifx \showURL      \undefined \def \showURL       {\relax}        \fi
\providecommand\bibfield[2]{#2}
\providecommand\bibinfo[2]{#2}
\providecommand\natexlab[1]{#1}
\providecommand\showeprint[2][]{arXiv:#2}

\bibitem[\protect\citeauthoryear{Acharya and Sidana}{Acharya and
  Sidana}{2007}]%
        {acharya2007classifying}
\bibfield{author}{\bibinfo{person}{Debashis Acharya} {and}
  \bibinfo{person}{Gajendra Sidana}.} \bibinfo{year}{2007}\natexlab{}.
\newblock \showarticletitle{Classifying mutual funds in India: Some results
  from clustering}.
\newblock \bibinfo{journal}{\emph{Indian Journal of Economics and Business}}
  \bibinfo{volume}{6}, \bibinfo{number}{1} (\bibinfo{year}{2007}),
  \bibinfo{pages}{71--79}.
\newblock


\bibitem[\protect\citeauthoryear{Aggarwal et~al\mbox{.}}{Aggarwal
  et~al\mbox{.}}{2016}]%
        {aggarwal2016recommender}
\bibfield{author}{\bibinfo{person}{Charu~C Aggarwal} {et~al\mbox{.}}}
  \bibinfo{year}{2016}\natexlab{}.
\newblock \bibinfo{booktitle}{\emph{Recommender systems}}.
\newblock \bibinfo{publisher}{Springer}.
\newblock


\bibitem[\protect\citeauthoryear{Agudo and L{\'a}zaro}{Agudo and
  L{\'a}zaro}{2005}]%
        {agudo2005does}
\bibfield{author}{\bibinfo{person}{Luis~Ferruz Agudo} {and}
  \bibinfo{person}{Cristina~Ortiz L{\'a}zaro}.}
  \bibinfo{year}{2005}\natexlab{}.
\newblock \showarticletitle{Does Mutual Fund Management in India correspond to
  its investment objective classification?}
\newblock \bibinfo{journal}{\emph{Review of Pacific Basin Financial Markets and
  Policies}} \bibinfo{volume}{8}, \bibinfo{number}{04} (\bibinfo{year}{2005}),
  \bibinfo{pages}{659--685}.
\newblock


\bibitem[\protect\citeauthoryear{Baghai-Wadji, El-Berry, Klocker, and
  Schwaiger}{Baghai-Wadji et~al\mbox{.}}{2005}]%
        {baghai2005consistency}
\bibfield{author}{\bibinfo{person}{Ramin Baghai-Wadji}, \bibinfo{person}{Rami
  El-Berry}, \bibinfo{person}{Stefan Klocker}, {and} \bibinfo{person}{Markus
  Schwaiger}.} \bibinfo{year}{2005}\natexlab{}.
\newblock \showarticletitle{The Consistency of Self-Declared Hedge Fund
  Styles—A Return-Based Analysis with Self-Organizing Maps}.
\newblock \bibinfo{journal}{\emph{Financial Stability Report}}
  \bibinfo{volume}{9} (\bibinfo{year}{2005}), \bibinfo{pages}{64--76}.
\newblock


\bibitem[\protect\citeauthoryear{Bengio, Goodfellow, and Courville}{Bengio
  et~al\mbox{.}}{2015}]%
        {bengio2015deep}
\bibfield{author}{\bibinfo{person}{Y. Bengio}, \bibinfo{person}{I.~J.
  Goodfellow}, {and} \bibinfo{person}{A. Courville}.}
  \bibinfo{year}{2015}\natexlab{}.
\newblock \showarticletitle{Deep learning}.
\newblock \bibinfo{journal}{\emph{MIT Press.}} (\bibinfo{year}{2015}).
\newblock


\bibitem[\protect\citeauthoryear{Breiman}{Breiman}{2001}]%
        {breiman2001random}
\bibfield{author}{\bibinfo{person}{Leo Breiman}.}
  \bibinfo{year}{2001}\natexlab{}.
\newblock \showarticletitle{Random forests}.
\newblock \bibinfo{journal}{\emph{Machine learning}} \bibinfo{volume}{45},
  \bibinfo{number}{1} (\bibinfo{year}{2001}), \bibinfo{pages}{5--32}.
\newblock


\bibitem[\protect\citeauthoryear{Brown and Goetzmann}{Brown and
  Goetzmann}{1997}]%
        {brown1997mutual}
\bibfield{author}{\bibinfo{person}{Stephen~J Brown} {and}
  \bibinfo{person}{William~N Goetzmann}.} \bibinfo{year}{1997}\natexlab{}.
\newblock \showarticletitle{Mutual fund styles}.
\newblock \bibinfo{journal}{\emph{Journal of financial Economics}}
  \bibinfo{volume}{43}, \bibinfo{number}{3} (\bibinfo{year}{1997}),
  \bibinfo{pages}{373--399}.
\newblock


\bibitem[\protect\citeauthoryear{Cai, Le-Khac, and Kechadi}{Cai
  et~al\mbox{.}}{2016}]%
        {cai2016clustering}
\bibfield{author}{\bibinfo{person}{Fan Cai}, \bibinfo{person}{Nhien-An
  Le-Khac}, {and} \bibinfo{person}{Tahar Kechadi}.}
  \bibinfo{year}{2016}\natexlab{}.
\newblock \showarticletitle{Clustering approaches for financial data analysis:
  a survey}.
\newblock \bibinfo{journal}{\emph{arXiv preprint arXiv:1609.08520}}
  (\bibinfo{year}{2016}).
\newblock


\bibitem[\protect\citeauthoryear{Castellanos and Alonso}{Castellanos and
  Alonso}{2005}]%
        {castellanos2005spanish}
\bibfield{author}{\bibinfo{person}{Arturo~Rodr{\'\i}guez Castellanos} {and}
  \bibinfo{person}{Bel{\'e}n~Vallejo Alonso}.} \bibinfo{year}{2005}\natexlab{}.
\newblock \showarticletitle{Spanish Mutual Fund Misclassification: Empirical
  Evidence}.
\newblock \bibinfo{journal}{\emph{The Journal of Investing}}
  \bibinfo{volume}{14}, \bibinfo{number}{1} (\bibinfo{year}{2005}),
  \bibinfo{pages}{41--51}.
\newblock


\bibitem[\protect\citeauthoryear{Chollet et~al\mbox{.}}{Chollet
  et~al\mbox{.}}{2015}]%
        {chollet2015keras}
\bibfield{author}{\bibinfo{person}{Fran\c{c}ois Chollet} {et~al\mbox{.}}}
  \bibinfo{year}{2015}\natexlab{}.
\newblock \bibinfo{title}{Keras}.
\newblock \bibinfo{howpublished}{\url{https://github.com/fchollet/keras}}.
\newblock


\bibitem[\protect\citeauthoryear{Corduas and Piccolo}{Corduas and
  Piccolo}{2008}]%
        {corduas2008time}
\bibfield{author}{\bibinfo{person}{Marcella Corduas} {and}
  \bibinfo{person}{Domenico Piccolo}.} \bibinfo{year}{2008}\natexlab{}.
\newblock \showarticletitle{Time series clustering and classification by the
  autoregressive metric}.
\newblock \bibinfo{journal}{\emph{Computational statistics \& data analysis}}
  \bibinfo{volume}{52}, \bibinfo{number}{4} (\bibinfo{year}{2008}),
  \bibinfo{pages}{1860--1872}.
\newblock


\bibitem[\protect\citeauthoryear{Cybenko}{Cybenko}{1989}]%
        {cybenko1989approximation}
\bibfield{author}{\bibinfo{person}{George Cybenko}.}
  \bibinfo{year}{1989}\natexlab{}.
\newblock \showarticletitle{Approximation by superpositions of a sigmoidal
  function}.
\newblock \bibinfo{journal}{\emph{Mathematics of control, signals and systems}}
  \bibinfo{volume}{2}, \bibinfo{number}{4} (\bibinfo{year}{1989}),
  \bibinfo{pages}{303--314}.
\newblock


\bibitem[\protect\citeauthoryear{Das et~al\mbox{.}}{Das et~al\mbox{.}}{2003}]%
        {das2003hedge}
\bibfield{author}{\bibinfo{person}{Nandita Das} {et~al\mbox{.}}}
  \bibinfo{year}{2003}\natexlab{}.
\newblock \showarticletitle{hedge Fund classification using K-means clustering
  Method}. In \bibinfo{booktitle}{\emph{9th International Conference on
  Computing in Economics and Finance}}. \bibinfo{pages}{11--13}.
\newblock


\bibitem[\protect\citeauthoryear{DiBartolomeo and Witkowski}{DiBartolomeo and
  Witkowski}{1997}]%
        {dibartolomeo1997mutual}
\bibfield{author}{\bibinfo{person}{Dan DiBartolomeo} {and}
  \bibinfo{person}{Erik Witkowski}.} \bibinfo{year}{1997}\natexlab{}.
\newblock \showarticletitle{Mutual fund misclassification: Evidence based on
  style analysis}.
\newblock \bibinfo{journal}{\emph{Financial Analysts Journal}}
  \bibinfo{volume}{53}, \bibinfo{number}{5} (\bibinfo{year}{1997}),
  \bibinfo{pages}{32--43}.
\newblock


\bibitem[\protect\citeauthoryear{Elton, Gruber, and Blake}{Elton
  et~al\mbox{.}}{2003}]%
        {elton2003incentive}
\bibfield{author}{\bibinfo{person}{Edwin~J Elton}, \bibinfo{person}{Martin~J
  Gruber}, {and} \bibinfo{person}{Christopher~R Blake}.}
  \bibinfo{year}{2003}\natexlab{}.
\newblock \showarticletitle{Incentive fees and mutual funds}.
\newblock \bibinfo{journal}{\emph{The Journal of Finance}}
  \bibinfo{volume}{58}, \bibinfo{number}{2} (\bibinfo{year}{2003}),
  \bibinfo{pages}{779--804}.
\newblock


\bibitem[\protect\citeauthoryear{Fawcett}{Fawcett}{2006}]%
        {10.1016/j.patrec.2005.10.010}
\bibfield{author}{\bibinfo{person}{Tom Fawcett}.}
  \bibinfo{year}{2006}\natexlab{}.
\newblock \showarticletitle{An Introduction to ROC Analysis}.
\newblock \bibinfo{journal}{\emph{Pattern Recogn. Lett.}} \bibinfo{volume}{27},
  \bibinfo{number}{8} (\bibinfo{date}{June} \bibinfo{year}{2006}),
  \bibinfo{pages}{861–874}.
\newblock
\showISSN{0167-8655}
\urldef\tempurl%
\url{https://doi.org/10.1016/j.patrec.2005.10.010}
\showDOI{\tempurl}


\bibitem[\protect\citeauthoryear{Gerlach and Maurer}{Gerlach and
  Maurer}{2017}]%
        {gerlach2017stable}
\bibfield{author}{\bibinfo{person}{Philipp Gerlach} {and}
  \bibinfo{person}{Raimond Maurer}.} \bibinfo{year}{2017}\natexlab{}.
\newblock \showarticletitle{Stable Return-Based Fund Classification}.
\newblock \bibinfo{journal}{\emph{Available at SSRN 2838187}}
  (\bibinfo{year}{2017}).
\newblock


\bibitem[\protect\citeauthoryear{Gibson and Gyger}{Gibson and Gyger}{2007}]%
        {gibson2007style}
\bibfield{author}{\bibinfo{person}{Rajna Gibson} {and}
  \bibinfo{person}{S{\'e}bastien Gyger}.} \bibinfo{year}{2007}\natexlab{}.
\newblock \showarticletitle{The style consistency of hedge funds}.
\newblock \bibinfo{journal}{\emph{European Financial Management}}
  \bibinfo{volume}{13}, \bibinfo{number}{2} (\bibinfo{year}{2007}),
  \bibinfo{pages}{287--308}.
\newblock


\bibitem[\protect\citeauthoryear{Hand and Till}{Hand and Till}{2001}]%
        {10.1023/A:1010920819831}
\bibfield{author}{\bibinfo{person}{David~J. Hand} {and}
  \bibinfo{person}{Robert~J. Till}.} \bibinfo{year}{2001}\natexlab{}.
\newblock \showarticletitle{A Simple Generalisation of the Area Under the ROC
  Curve for Multiple Class Classification Problems}.
\newblock \bibinfo{journal}{\emph{Mach. Learn.}} \bibinfo{volume}{45},
  \bibinfo{number}{2} (\bibinfo{date}{Oct.} \bibinfo{year}{2001}),
  \bibinfo{pages}{171–186}.
\newblock
\showISSN{0885-6125}
\urldef\tempurl%
\url{https://doi.org/10.1023/A:1010920819831}
\showDOI{\tempurl}


\bibitem[\protect\citeauthoryear{Haslem and Scheraga}{Haslem and
  Scheraga}{2001}]%
        {haslem2001morningstar}
\bibfield{author}{\bibinfo{person}{John~A Haslem} {and} \bibinfo{person}{Carl~A
  Scheraga}.} \bibinfo{year}{2001}\natexlab{}.
\newblock \showarticletitle{Morningstar's classification of large-cap mutual
  funds}.
\newblock \bibinfo{journal}{\emph{The Journal of Investing}}
  \bibinfo{volume}{10}, \bibinfo{number}{1} (\bibinfo{year}{2001}),
  \bibinfo{pages}{79--89}.
\newblock


\bibitem[\protect\citeauthoryear{Hastie, Tibshirani, and Friedman}{Hastie
  et~al\mbox{.}}{2009}]%
        {hastie2009elements}
\bibfield{author}{\bibinfo{person}{Trevor Hastie}, \bibinfo{person}{Robert
  Tibshirani}, {and} \bibinfo{person}{Jerome Friedman}.}
  \bibinfo{year}{2009}\natexlab{}.
\newblock \bibinfo{booktitle}{\emph{The elements of statistical learning: data
  mining, inference, and prediction}}.
\newblock \bibinfo{publisher}{Springer Science \& Business Media}.
\newblock


\bibitem[\protect\citeauthoryear{Hornik, Stinchcombe, and White}{Hornik
  et~al\mbox{.}}{1989}]%
        {hornik1989multilayer}
\bibfield{author}{\bibinfo{person}{Kurt Hornik}, \bibinfo{person}{Maxwell
  Stinchcombe}, {and} \bibinfo{person}{Halbert White}.}
  \bibinfo{year}{1989}\natexlab{}.
\newblock \showarticletitle{Multilayer feedforward networks are universal
  approximators}.
\newblock \bibinfo{journal}{\emph{Neural networks}} \bibinfo{volume}{2},
  \bibinfo{number}{5} (\bibinfo{year}{1989}), \bibinfo{pages}{359--366}.
\newblock


\bibitem[\protect\citeauthoryear{Kim, Shukla, and Tomas}{Kim
  et~al\mbox{.}}{2000}]%
        {kim2000mutual}
\bibfield{author}{\bibinfo{person}{Moon Kim}, \bibinfo{person}{Ravi Shukla},
  {and} \bibinfo{person}{Michael Tomas}.} \bibinfo{year}{2000}\natexlab{}.
\newblock \showarticletitle{Mutual fund objective misclassification}.
\newblock \bibinfo{journal}{\emph{Journal of Economics and Business}}
  \bibinfo{volume}{52}, \bibinfo{number}{4} (\bibinfo{year}{2000}),
  \bibinfo{pages}{309--323}.
\newblock


\bibitem[\protect\citeauthoryear{Kingma and Ba}{Kingma and Ba}{2014}]%
        {kingma2014adam}
\bibfield{author}{\bibinfo{person}{Diederik~P Kingma} {and}
  \bibinfo{person}{Jimmy Ba}.} \bibinfo{year}{2014}\natexlab{}.
\newblock \showarticletitle{Adam: A method for stochastic optimization}.
\newblock \bibinfo{journal}{\emph{arXiv preprint arXiv:1412.6980}}
  (\bibinfo{year}{2014}).
\newblock


\bibitem[\protect\citeauthoryear{Kohonen}{Kohonen}{1990}]%
        {kohonen1990self}
\bibfield{author}{\bibinfo{person}{Teuvo Kohonen}.}
  \bibinfo{year}{1990}\natexlab{}.
\newblock \showarticletitle{The self-organizing map}.
\newblock \bibinfo{journal}{\emph{Proc. IEEE}} \bibinfo{volume}{78},
  \bibinfo{number}{9} (\bibinfo{year}{1990}), \bibinfo{pages}{1464--1480}.
\newblock


\bibitem[\protect\citeauthoryear{Lajbcygier and Yahya}{Lajbcygier and
  Yahya}{2008}]%
        {lajbcygier2008soft}
\bibfield{author}{\bibinfo{person}{Paul Lajbcygier} {and}
  \bibinfo{person}{Asjad Yahya}.} \bibinfo{year}{2008}\natexlab{}.
\newblock \showarticletitle{Soft Clustering for Funds Management Style
  Analysis: Out-of-Sample Predictability}.
\newblock \bibinfo{journal}{\emph{Available at SSRN 1206731}}
  (\bibinfo{year}{2008}).
\newblock


\bibitem[\protect\citeauthoryear{Lamponi}{Lamponi}{2015}]%
        {lamponi2015data}
\bibfield{author}{\bibinfo{person}{Daniele Lamponi}.}
  \bibinfo{year}{2015}\natexlab{}.
\newblock \showarticletitle{A Data-Driven Categorization of Investable Assets}.
\newblock \bibinfo{journal}{\emph{The Journal of Investing}}
  \bibinfo{volume}{24}, \bibinfo{number}{4} (\bibinfo{year}{2015}),
  \bibinfo{pages}{73--80}.
\newblock


\bibitem[\protect\citeauthoryear{LeCun, Bengio, and Hinton}{LeCun
  et~al\mbox{.}}{2015}]%
        {lecun2015deep}
\bibfield{author}{\bibinfo{person}{Y. LeCun}, \bibinfo{person}{Y. Bengio},
  {and} \bibinfo{person}{G. Hinton}.} \bibinfo{year}{2015}\natexlab{}.
\newblock \showarticletitle{Deep learning}.
\newblock \bibinfo{journal}{\emph{Nature}} \bibinfo{volume}{521},
  \bibinfo{number}{7553} (\bibinfo{year}{2015}), \bibinfo{pages}{436--444}.
\newblock


\bibitem[\protect\citeauthoryear{Liaw, Wiener, et~al\mbox{.}}{Liaw
  et~al\mbox{.}}{2002}]%
        {liaw2002classification}
\bibfield{author}{\bibinfo{person}{Andy Liaw}, \bibinfo{person}{Matthew
  Wiener}, {et~al\mbox{.}}} \bibinfo{year}{2002}\natexlab{}.
\newblock \showarticletitle{Classification and regression by randomForest}.
\newblock \bibinfo{journal}{\emph{R news}} \bibinfo{volume}{2},
  \bibinfo{number}{3} (\bibinfo{year}{2002}), \bibinfo{pages}{18--22}.
\newblock


\bibitem[\protect\citeauthoryear{lipper category}{lipper category}{[n.d.]}]%
        {lippercategory}
lipper category \bibinfo{year}{[n.d.]}\natexlab{}.
\newblock \bibinfo{title}{Rifinitive. "Lipper Fund Research."}.
\newblock
  \bibinfo{howpublished}{\href{https://www.refinitiv.com/en/products/lipper-fund-research}}.
\newblock


\bibitem[\protect\citeauthoryear{Lisi and Otranto}{Lisi and Otranto}{2010}]%
        {lisi2010clustering}
\bibfield{author}{\bibinfo{person}{Francesco Lisi} {and}
  \bibinfo{person}{Edoardo Otranto}.} \bibinfo{year}{2010}\natexlab{}.
\newblock \showarticletitle{Clustering mutual funds by return and risk levels}.
\newblock In \bibinfo{booktitle}{\emph{Mathematical and statistical methods for
  actuarial sciences and finance}}. \bibinfo{publisher}{Springer},
  \bibinfo{pages}{183--191}.
\newblock


\bibitem[\protect\citeauthoryear{Lytkin, Kulikowski, and Muchnik}{Lytkin
  et~al\mbox{.}}{2008}]%
        {lytkin2008variance}
\bibfield{author}{\bibinfo{person}{Nikita~I Lytkin}, \bibinfo{person}{Casimir~A
  Kulikowski}, {and} \bibinfo{person}{Ilya~B Muchnik}.}
  \bibinfo{year}{2008}\natexlab{}.
\newblock \bibinfo{booktitle}{\emph{Variance-based criteria for clustering and
  their application to the analysis of management styles of mutual funds based
  on time series of daily returns}}.
\newblock \bibinfo{type}{{T}echnical {R}eport}. \bibinfo{institution}{DIMACS
  Technical Report 2008-01}.
\newblock


\bibitem[\protect\citeauthoryear{Marathe and Shawky}{Marathe and
  Shawky}{1999}]%
        {marathe1999categorizing}
\bibfield{author}{\bibinfo{person}{Achla Marathe} {and} \bibinfo{person}{Hany~A
  Shawky}.} \bibinfo{year}{1999}\natexlab{}.
\newblock \showarticletitle{Categorizing mutual funds using clusters}.
\newblock \bibinfo{journal}{\emph{Advances in Quantitative analysis of Finance
  and Accounting}} \bibinfo{volume}{7}, \bibinfo{number}{1}
  (\bibinfo{year}{1999}), \bibinfo{pages}{199--204}.
\newblock


\bibitem[\protect\citeauthoryear{Martin, Keown~Jr, and Farrell}{Martin
  et~al\mbox{.}}{1982}]%
        {martin1982fund}
\bibfield{author}{\bibinfo{person}{John~D Martin}, \bibinfo{person}{Arthur~J
  Keown~Jr}, {and} \bibinfo{person}{James~L Farrell}.}
  \bibinfo{year}{1982}\natexlab{}.
\newblock \showarticletitle{Do fund objectives affect diversification
  policies?}
\newblock \bibinfo{journal}{\emph{Journal of Portfolio Management}}
  \bibinfo{volume}{8}, \bibinfo{number}{2} (\bibinfo{year}{1982}),
  \bibinfo{pages}{19--28}.
\newblock


\bibitem[\protect\citeauthoryear{McDonald}{McDonald}{1974}]%
        {mcdonald1974objectives}
\bibfield{author}{\bibinfo{person}{John~G McDonald}.}
  \bibinfo{year}{1974}\natexlab{}.
\newblock \showarticletitle{Objectives and performance of mutual funds,
  1960--1969}.
\newblock \bibinfo{journal}{\emph{Journal of Financial and Quantitative
  Analysis}} \bibinfo{volume}{9}, \bibinfo{number}{3} (\bibinfo{year}{1974}),
  \bibinfo{pages}{311--333}.
\newblock


\bibitem[\protect\citeauthoryear{Menardi and Lisi}{Menardi and Lisi}{2015}]%
        {menardi2015double}
\bibfield{author}{\bibinfo{person}{Giovanna Menardi} {and}
  \bibinfo{person}{Francesco Lisi}.} \bibinfo{year}{2015}\natexlab{}.
\newblock \showarticletitle{Double clustering for rating mutual funds}.
\newblock \bibinfo{journal}{\emph{Electronic Journal of Applied Statistical
  Analysis}} \bibinfo{volume}{8}, \bibinfo{number}{1} (\bibinfo{year}{2015}),
  \bibinfo{pages}{44--56}.
\newblock


\bibitem[\protect\citeauthoryear{Miceli and Susinno}{Miceli and
  Susinno}{2004}]%
        {miceli2004ultrametricity}
\bibfield{author}{\bibinfo{person}{Maria-Augusta Miceli} {and}
  \bibinfo{person}{Gabriele Susinno}.} \bibinfo{year}{2004}\natexlab{}.
\newblock \showarticletitle{Ultrametricity in fund of funds diversification}.
\newblock \bibinfo{journal}{\emph{Physica A: Statistical Mechanics and its
  Applications}} \bibinfo{volume}{344}, \bibinfo{number}{1-2}
  (\bibinfo{year}{2004}), \bibinfo{pages}{95--99}.
\newblock


\bibitem[\protect\citeauthoryear{Moreno, Marco, and Olmeda}{Moreno
  et~al\mbox{.}}{2006}]%
        {moreno2006self}
\bibfield{author}{\bibinfo{person}{David Moreno}, \bibinfo{person}{Paulina
  Marco}, {and} \bibinfo{person}{Ignacio Olmeda}.}
  \bibinfo{year}{2006}\natexlab{}.
\newblock \showarticletitle{Self-organizing maps could improve the
  classification of Spanish mutual funds}.
\newblock \bibinfo{journal}{\emph{European Journal of Operational Research}}
  \bibinfo{volume}{174}, \bibinfo{number}{2} (\bibinfo{year}{2006}),
  \bibinfo{pages}{1039--1054}.
\newblock


\bibitem[\protect\citeauthoryear{Morningstar Gloabl Equity}{Morningstar Gloabl
  Equity}{2019}]%
        {morningstarequitysectorbreakdown}
Morningstar Gloabl Equity \bibinfo{year}{2019}\natexlab{}.
\newblock \bibinfo{title}{Morningstar Gloabl Equity Classification Structure}.
\newblock
  \bibinfo{howpublished}{\href{https://indexes.morningstar.com/resources/PDF/Methodology\%20Documents/SectorArticle.pdf}{Equity
  Classification Structure}}.
\newblock


\bibitem[\protect\citeauthoryear{Morningstar Global Category}{Morningstar
  Global Category}{[n.d.]}]%
        {morningstarglobalcategory}
Morningstar Global Category \bibinfo{year}{[n.d.]}\natexlab{}.
\newblock \bibinfo{title}{Morningstar $Global Category^{TM}$ Classifications}.
\newblock
  \bibinfo{howpublished}{\href{https://www.morningstar.com/content/dam/marketing/shared/research/methodology/860250-GlobalCategoryClassifications.pdf}{GlobalCategoryClassifications}}.
\newblock


\bibitem[\protect\citeauthoryear{Morningstar Global Fixed Income}{Morningstar
  Global Fixed Income}{2017}]%
        {morningstarfixedincomeclassificationreakdown}
Morningstar Global Fixed Income \bibinfo{year}{2017}\natexlab{}.
\newblock \bibinfo{title}{Morningstar Global Fixed Income Classification}.
\newblock
  \bibinfo{howpublished}{\href{https://www.morningstar.com/content/dam/marketing/shared/research/methodology/829856-Morningstar_Global_Fixed_Income_Classification_Methodology.pdf}{Fixed
  Income Classification}}.
\newblock


\bibitem[\protect\citeauthoryear{Orphanides et~al\mbox{.}}{Orphanides
  et~al\mbox{.}}{1996}]%
        {orphanides1996compensation}
\bibfield{author}{\bibinfo{person}{Athanasios Orphanides} {et~al\mbox{.}}}
  \bibinfo{year}{1996}\natexlab{}.
\newblock \bibinfo{booktitle}{\emph{Compensation incentives and risk taking
  behavior: evidence from mutual funds}}.
\newblock \bibinfo{publisher}{Citeseer}.
\newblock


\bibitem[\protect\citeauthoryear{Pattarin, Paterlini, and Minerva}{Pattarin
  et~al\mbox{.}}{2004}]%
        {pattarin2004clustering}
\bibfield{author}{\bibinfo{person}{Francesco Pattarin}, \bibinfo{person}{Sandra
  Paterlini}, {and} \bibinfo{person}{Tommaso Minerva}.}
  \bibinfo{year}{2004}\natexlab{}.
\newblock \showarticletitle{Clustering financial time series: an application to
  mutual funds style analysis}.
\newblock \bibinfo{journal}{\emph{Computational Statistics \& Data Analysis}}
  \bibinfo{volume}{47}, \bibinfo{number}{2} (\bibinfo{year}{2004}),
  \bibinfo{pages}{353--372}.
\newblock


\bibitem[\protect\citeauthoryear{Pedregosa, Varoquaux, Gramfort, Michel,
  Thirion, Grisel, Blondel, Prettenhofer, Weiss, Dubourg, Vanderplas, Passos,
  Cournapeau, Brucher, Perrot, and Duchesnay}{Pedregosa et~al\mbox{.}}{2011}]%
        {scikit-learn}
\bibfield{author}{\bibinfo{person}{F. Pedregosa}, \bibinfo{person}{G.
  Varoquaux}, \bibinfo{person}{A. Gramfort}, \bibinfo{person}{V. Michel},
  \bibinfo{person}{B. Thirion}, \bibinfo{person}{O. Grisel},
  \bibinfo{person}{M. Blondel}, \bibinfo{person}{P. Prettenhofer},
  \bibinfo{person}{R. Weiss}, \bibinfo{person}{V. Dubourg}, \bibinfo{person}{J.
  Vanderplas}, \bibinfo{person}{A. Passos}, \bibinfo{person}{D. Cournapeau},
  \bibinfo{person}{M. Brucher}, \bibinfo{person}{M. Perrot}, {and}
  \bibinfo{person}{E. Duchesnay}.} \bibinfo{year}{2011}\natexlab{}.
\newblock \showarticletitle{Scikit-learn: Machine Learning in {P}ython}.
\newblock \bibinfo{journal}{\emph{Journal of Machine Learning Research}}
  \bibinfo{volume}{12} (\bibinfo{year}{2011}), \bibinfo{pages}{2825--2830}.
\newblock


\bibitem[\protect\citeauthoryear{Resnick and Varian}{Resnick and
  Varian}{1997}]%
        {resnick1997recommender}
\bibfield{author}{\bibinfo{person}{Paul Resnick} {and} \bibinfo{person}{Hal~R
  Varian}.} \bibinfo{year}{1997}\natexlab{}.
\newblock \showarticletitle{Recommender systems}.
\newblock \bibinfo{journal}{\emph{Commun. ACM}} \bibinfo{volume}{40},
  \bibinfo{number}{3} (\bibinfo{year}{1997}), \bibinfo{pages}{56--59}.
\newblock


\bibitem[\protect\citeauthoryear{Sakakibara, Matsui, Mutoh, and
  Inuzuka}{Sakakibara et~al\mbox{.}}{2015}]%
        {sakakibara2015clustering}
\bibfield{author}{\bibinfo{person}{Takumasa Sakakibara},
  \bibinfo{person}{Tohgoroh Matsui}, \bibinfo{person}{Atsuko Mutoh}, {and}
  \bibinfo{person}{Nobuhiro Inuzuka}.} \bibinfo{year}{2015}\natexlab{}.
\newblock \showarticletitle{Clustering mutual funds based on investment
  similarity}.
\newblock \bibinfo{journal}{\emph{Procedia Computer Science}}
  \bibinfo{volume}{60} (\bibinfo{year}{2015}), \bibinfo{pages}{881--890}.
\newblock


\bibitem[\protect\citeauthoryear{Sharpe}{Sharpe}{1992}]%
        {sharpe1992asset}
\bibfield{author}{\bibinfo{person}{William~F Sharpe}.}
  \bibinfo{year}{1992}\natexlab{}.
\newblock \showarticletitle{Asset allocation: Management style and performance
  measurement}.
\newblock \bibinfo{journal}{\emph{Journal of portfolio Management}}
  \bibinfo{volume}{18}, \bibinfo{number}{2} (\bibinfo{year}{1992}),
  \bibinfo{pages}{7--19}.
\newblock


\bibitem[\protect\citeauthoryear{Shawky and Marathe}{Shawky and
  Marathe}{2010}]%
        {shawky2010stylistic}
\bibfield{author}{\bibinfo{person}{Hany~A Shawky} {and} \bibinfo{person}{Achla
  Marathe}.} \bibinfo{year}{2010}\natexlab{}.
\newblock \showarticletitle{Stylistic Differences across Hedge Funds as
  Revealed by Historical Monthly Returns}.
\newblock \bibinfo{journal}{\emph{Technology and Investment}}
  \bibinfo{volume}{1}, \bibinfo{number}{01} (\bibinfo{year}{2010}),
  \bibinfo{pages}{26}.
\newblock


\bibitem[\protect\citeauthoryear{Vozlyublennaia and Wu}{Vozlyublennaia and
  Wu}{2018}]%
        {vozlyublennaia2018mutual}
\bibfield{author}{\bibinfo{person}{Nadia Vozlyublennaia} {and}
  \bibinfo{person}{Youchang Wu}.} \bibinfo{year}{2018}\natexlab{}.
\newblock \showarticletitle{Mutual funds apart from the crowd}.
\newblock \bibinfo{journal}{\emph{Available at SSRN 2769161}}
  (\bibinfo{year}{2018}).
\newblock


\end{thebibliography}



\end{document}